# On the weight and density bounds of Polynomial threshold functions

Erhan Oztop[1,2], Minoru Asada[1]

[1]Osaka University, Japan, [2]Ozyegin University, Turkey

**Abstract.** In this report, we show that all n-variable Boolean function can be represented as polynomial threshold functions (PTF) with at most $0.75 \times 2^n$ non-zero integer coefficients and give an upper bound on the absolute value of these coefficients. To our knowledge this provides the best known bound on both the PTF density (number of monomials) and weight (sum of the coefficient magnitudes) of general Boolean functions. The special case of Bent functions is also analyzed and shown that any n-variable Bent function can be represented with integer coefficients less than $2^n$ while also obeying the aforementioned density bound. Finally, sparse Boolean functions, which are almost constant except for $m \ll 2^n$ number of variable assignments, are shown to have small weight PTFs with density at most $m + 2^{n-1}$.

## 1. Introduction

We consider Boolean functions $f: \{-1,1\}^n \to \{-1,1\}$ and study their polynomial threshold function (PTF) representations in terms of the number of monomials and the magnitude of the polynomial coefficients. PTF representation is also called polynomial sign-representation, and in the literature, abstract units computing PTFs appear under the names of PTF units, higher-order neurons, product units, or sigma-pi units [1-3]. As it is known that the use of higher-order units increases the computational power and storage capacities of neural networks [4-6], past research in neural networks have focused on developing algorithms for finding a small set of monomials to sign-represent a given Boolean function (BF) without suffering from the combinatorial growth problem [e.g. 6, 7-9]. Besides the practical application of PTFs, it is of theoretical importance to know the extremal properties of PTF representation of Boolean functions. The theoretical studies on this front come from extremal combinatorics [e.g. 10, 11, 12] and circuit complexity [e.g. 13, 14]. The three PTF complexity measures focused in the literature are *degree*, *density* and *weight* of BFs [10, 15]. The PTF *degree* of a BF $f$, refers to the minimum degree polynomial that can sign-represent f. The PTF *density* of $f$ is defined as the minimum number of monomials with which one can sign-represent $f$. Finally, the PTF weight of $f$ is the smallest sum of the absolute values of the weights in an integer-weight PTF representation of $f$. This paper is focused on the latter two measures.

In an earlier report we have derived a non-asymptotic upper bound, $0.75 \times 2^n$ on the PTF density of n-variable BFs, which, to our knowledge, is still the best upper bound known [16]. In this report, we show that it is possible to obtain integer-weight PTF representation with the same density bound for any BF and derive an upper bound on the weight of such representation. We then direct our attention to Bent functions and prove that they assume integer coefficient PTF representations with surprisingly low absolute values, i.e. at most $2^n$, while still having a density of at most. Finally, we derive PTF density and weight upper bounds for sparse Boolean functions, which are almost constant except for $m \ll 2^n$ number of variable assignments.

## 2. Basic definitions and results

We use bold lower-case letters to indicate vectors and bold upper case for matrices. The vectors are column vectors unless otherwise stated. Table 1 lists the acronyms and notations used in the paper. In some cases, the precise meaning of the terms used in the table will become clear with their full definition given subsequently in the text.

*Table 1. The acronyms and notations used in the paper.*

| | |
|---|---|
| BF | Boolean Function |
| PTF | Polynomial Threshold Function |
| $\boldsymbol{y}$ | A real column vector |
| $\boldsymbol{Y}$ | A real matrix |
| $\boldsymbol{I}$ | The identity matrix (sometimes size is given as a subscript) |
| $\mathcal{H}_n$ | Sylvester type Hadamard matrix of order $2^n$ |
| $\mathcal{H}$ | A Sylvester type Hadamard matrix (when order is clear from the text) |
| $\lvert\boldsymbol{y}\rvert$ | Component wise absolute value of the real vector $\boldsymbol{y}$ |
| $\lceil\boldsymbol{y}\rceil$ | The maximum component in $\lvert\boldsymbol{y}\rvert$ |
| $\text{sgn}(\boldsymbol{y})$ | The sign function applied to each component of $\boldsymbol{y}$ |
| $\text{diag}(\boldsymbol{y})$ | The square matrix where the diagonal elements are taken from $\boldsymbol{y}$ |
| $\log(x)$ | The base-2 logarithm of $x$ |
| $\boldsymbol{1}$ | All ones column vector |
| $\boldsymbol{0}$ | All zeros column vector |
| $\boldsymbol{f}$ | If $f$ is a n-variable BF, then $\boldsymbol{f}$ is its $2^n$ component vector representation |
| $\#\boldsymbol{A}$ | The number of rows in the matrix $\boldsymbol{A}$ |
| $\mathcal{D}_f$ | PTF density of $f$ |
| $\mathcal{W}_f$ | PTF weight of $f$ |

In this study, we consider Boolean functions over the domain $\{-1,1\}^n$ where $-1$ and $1$ corresponds to binary variables of 1 (True) and False (0) respectively. A binary variable $b \in \{0,1\}$ can be converted to $\{-1,1\}$ domain with the transformation $(-1)^b$. By adopting a fixed ordering for the truth assignments of the input variables, an n-variable Boolean function $f$ can be represented as $2^n$ sized $\pm 1$ vector $\boldsymbol{f}$. A BF have several other useful representations. In particular, in this report, the coefficients of multi-linear polynomials either interpolating or matching the sign of $f$ at every variable assignment will be utilized. We describe these representations concretely in the following definitions.

*Definition* (Spectrum). Any n-variable Boolean function $f$ has a unique representation as the weighted sum of monomials $f(x_1, x_2, \cdots, x_n) = \sum_{i=1}^{2^n} s_i \prod_{k \in S_i} x_k$ where $S_i \subset \{1, 2, \ldots, n\}$. This can be seen for example by direct application of Lagrange interpolation. The coefficient vector **s** is called the spectrum, or the Fourier coefficients of $f$ denoted by $\hat{f}$.

*Definition* (Sign Representation/Polynomial Threshold Function). A multilinear polynomial $p$ is said to *sign-represent* an n-variable Boolean function $f$ if $f(x_1, x_2, \cdots, x_n) = \text{sgn}(p(x_1, x_2, \cdots, x_n))$ for all assignments $[x_1, x_2, \cdots, x_n]^T \in \{-1, 1\}^n$. In this case we say $p$ is a Polynomial Threshold Function (PTF) representation of $f$, or $p$ sign-represents f.

*Definition* (Walsh spectrum, Walsh polynomial). The $2^n$ scaled version of the exact interpolating polynomial of BF $f$ is called the Walsh polynomial of $f$. Consequently, the spectrum scaled by $2^n$ is called the Walsh spectrum or Walsh coefficients of $f$.

*Definition* (Sylvester type Hadamard matrix). A Sylvester type Hadamard matrix of order $2^n$ ($n > 0$) can be defined as [see e.g. 13, 17]

$$\mathcal{H}_n = \begin{cases} [1] & \text{if } n = 0 \\ \begin{bmatrix} \mathcal{H}_{n-1} & \mathcal{H}_{n-1} \\ \mathcal{H}_{n-1} & -\mathcal{H}_{n-1} \end{bmatrix} & \text{if } n > 0 \end{cases} \qquad \text{Eq. 1}$$

**Lemma 1**. *A Sylvester type Hadamard matrix $\mathcal{H}_n$ is symmetric, orthogonal and has $2^{-n}\mathcal{H}_n$ as the inverse.*

*Proof*. Symmetricity is evident from the definition. Orthogonality and the expression for the inverse can be obtained by applying induction on the following identity.

$$\mathcal{H}_n \mathcal{H}_n^T = \begin{cases} 1 & \text{if } n = 0 \\ 2\begin{bmatrix} \mathcal{H}_{n-1}\mathcal{H}_{n-1}^T & \mathbf{0} \\ \mathbf{0} & \mathcal{H}_{n-1}\mathcal{H}_{n-1}^T \end{bmatrix} & \text{if } n > 0 \end{cases}$$

□

*Definition* (Exact polynomial representation in vector form). One can choose a natural ordering for the truth assignments of the variables and the monomials such that the condition $f(x_1, x_2, \cdots, x_n) = p(x_1, x_2, \cdots, x_n)$ can be precisely expressed with $\boldsymbol{f} = \mathcal{H}\boldsymbol{s}$ where $\mathcal{H}$ is a $2^n \times 2^n$ Sylvester type Hadamard matrix [13, 16]. Note that here $\boldsymbol{f}$ is the vector representation of $f$ while $\boldsymbol{s}$ is its spectrum.

**Lemma 2**. *The spectrum, $\boldsymbol{s}$ of the polynomial representation of a Boolean function, $\boldsymbol{f}$ is given by $\boldsymbol{s} = 2^{-n}\mathcal{H}_n\boldsymbol{f}$. Consequently, the Walsh spectrum is given by $\boldsymbol{\varpi} = \mathcal{H}_n\boldsymbol{f}$*

*Proof*. Exact interpolation at each assignment means $\boldsymbol{f} = \mathcal{H}_n\boldsymbol{s}$. Multiplying both sides with the inverse of $\mathcal{H}_n$ gives the desired result. □

**Lemma 3**. *The norm of the spectrum ($s$) of any BF is 1. Consequently, the norm of the Walsh spectrum ($\varpi$) of any BF is $2^n$.*

*Proof.* Use Lemma 2 for the expression of the spectrum $s$ and compute $\|s\|^2$:
$\|s\|^2 = (2^{-n}\mathcal{H}\mathbf{f})^T \, 2^{-n}\mathcal{H}\mathbf{f} = 2^{-2n}\mathbf{f}^T\mathcal{H}\mathcal{H}\mathbf{f} = 2^{-n}\mathbf{f}^T\mathbf{f} = 1$ which says $\|s\| = 1$ and $\|\varpi\| = 2^n$. □

*Definition* (The vector form of sign-representation). Using the natural ordering introduced above, the sign-representation condition can be expressed as $Y\mathcal{H}_n w > 0$ where $Y = \text{diag}(f)$, which is called the vector form of sign-representation.

By using the vector form of sign-representation it is easy to see that there is a one-to-one correspondence between sign representations and the positive half space of $\mathbb{R}^{2^n}$ due to the orthogonality of $\mathcal{H}_n$. Below we state this as a theorem.

**Theorem 1**. *For a given BF $f$, Let $Y = diag(f)$ then $\mathcal{H}_n w > 0$ if and only if $w \in \{\mathcal{H}_n Y k \mid k \in \mathbb{R}^{2^n} \text{ and } k > 0\}$.*

*Proof.* $Y\mathcal{H}_n w > 0 \Leftrightarrow Y\mathcal{H}_n w = k$ for some $k \in \mathbb{R}^{2^n}$ with $k > 0$. Multiplying the equation with the inverse of $Y\mathcal{H}_n$ gives the desired result. □

## 3. Contributions

To present the contributions of this study concretely we make the following definitions.

*Definition (*Density/length*).* The length or density of a PTF is the number of non-zero weight input lines. Equivalently, it is the number of monomials in a given PTF representation of a BF.

*Definition (*Weight*).* The weight of a PTF (representing some BF $f$) is the sum of the absolute values of the coefficients.

*Definition* (PTF density, $\mathcal{D}_f$). PTF density of a BF $f$ ($\mathcal{D}_f$) is the minimum number of monomials that one can sign-represent it. Equivalently, density of $f$ is the minimum length PTF representation of it. When the function is clear from the context or when all BFs are implied then subscript $f$ may be dropped.

*Definition* (PTF weight, $\mathcal{W}_f$). PTF weight of a BF $f$ ($\mathcal{W}_f$) is the minimum sum of the absolute values of the coefficients over all possible PTF representations of $f$ with integer coefficients. When we wish to constrain the weight to set of PTFs satisfying a specific condition, we write the condition expression in the square brackets, as in $\mathcal{W}_f[expression]$. When the function is clear from the context or when all BFs are implied then subscript $f$ may be dropped.

We can now concisely state the contributions of the current study and give the related results (Table 2):

- Analyze the proof [16] that establishes $\mathcal{D}_f \leq 0.75 \times 2^n$ for all $f$
- Derive an upper bound on $\mathcal{W}_f[\mathcal{D}_f \leq 0.75 \times 2^n]$ over all BF $f$.
- Derive an upper bound on $\mathcal{W}_{f_{bent}}[\mathcal{D}_{f_{bent}} \leq 0.75 \times 2^n]$ over all Bent functions
- Derive an upper bound on $\mathcal{D}_{f_{sp}}$ over all m-sparse functions $f_{sp}$
- Derive an upper bound on $\mathcal{W}_{f_{sp}}[\mathcal{D}_{f_{sp}} \leq m + 2^{n-1}]$ over all m-sparse functions $f_{sp}$

*Table 2. The summary of the results presented/obtained. $\mathcal{D}$ and $\mathcal{W}$ represents PTF density and PTF weight respectively.*

| PTF measure | Function class | Upper bound | Source |
|---|---|---|---|
| $\mathcal{D}_f$ | All BFs | $0.75 \times 2^n$ | Theorem 2, (Oztop[16]) |
| $\mathcal{W}_f[\mathcal{D}_f \leq 0.75 \times 2^n]$ | All BFs | $0.75 \times 2^{n2^{n-3}-2^{n-1}+5n}$ | Theorem 3, corollary |
| $\mathcal{W}_{f_{bent}}[\mathcal{D}_{f_{bent}} \leq 0.75 \times 2^n]$ | Bent functions | $0.75 \times 2^{2n}$ | Theorem 4, corollary |
| $\mathcal{D}_{f_{sp}}$ | m-sparse functions | $2^{n-1} + min(m, 2^{n-2})$ | Theorem 5 |
| $\mathcal{W}_{f_{sp}}[\mathcal{D}_{f_{sp}} \leq m + 2^{n-1}]$ | m-sparse functions | $3 \times 2^{2n+0.5m \log m - m + 1.5 \log m}$ | Theorem 6 corollary |

The PTF weight bounds derived in this report utilizes the well-known Hadamard inequality together with the logic of the proof followed in our earlier work [16]. We state the former as a lemma without proof and give a sketch of the proof for the latter in the next section.

**Lemma 4.** *(Hadamard). If $\mathbf{A}$ is an $n \times n$ matrix, then $|det\,\mathbf{A}| \leq \prod_{i=1}^{n}\left(\sum_{j=1}^{n} A_{ij}^2\right)^{1/2}$ where $A_{ij}$ is the matrix entry at row i and column j of $\mathbf{A}$. Furthermore, the equality is attained if and only if $\mathbf{A}$ is an orthogonal matrix.*

A simple but useful result we borrow for weight bound estimations is due to Håstad on the divisibility of the determinants of $\pm 1$ matrices [18]:

**Lemma 5** *(Håstad). The determinant of an $m \times m$ matrix $\mathbf{A}$ with entries $\pm 1$ is divisible by $2^{m-1}$.*

*Proof.* Adding row 1 to the other rows in $\mathbf{A}$ produce a new matrix $\widetilde{\mathbf{A}}$ where rows 2,3,...m have only zeros and $\pm 2$ entries. Since adding a scaled version of a row to another row does not change the determinant $det\,\mathbf{A} = det\,\widetilde{\mathbf{A}}$. Factoring out 2's from $det\,\widetilde{\mathbf{A}}$ shows that $det\,\widetilde{\mathbf{A}} = det\,\mathbf{A}$ is divisible by $2^{m-1}$ □

## 4. PTF representation with length at most $0.75 \times 2^n$

We use the recipe given in [16] to construct PTFs with $0.75 \times 2^n$ or smaller number of monomials; but while doing so we ensure that the weights found are integers, and then bound these integers from above. Therefore, a review of the theorem given in [16] is in order.

***Theorem 2*** *(Oztop). For any n-variable Boolean function, there exists a sign-representing polynomial with $0.75 \times 2^n$ or a smaller number of monomials.*

*Proof.* (Sketch) Remembering that all the PTF representations of a Boolean function $\mathbf{y}$ are characterized by the inequality system $\text{diag}(\mathbf{y})\mathcal{H}_n\mathbf{w} > \mathbf{0}$, we represent the upper and lower halves of the coefficient vector $\mathbf{w}$ with $\mathbf{u}$ and $\mathbf{v}$, and expand $\mathcal{H}_n$ by noting that $\mathcal{H}_n = \begin{bmatrix} \mathcal{H}_{n-1} & \mathcal{H}_{n-1} \\ \mathcal{H}_{n-1} & -\mathcal{H}_{n-1} \end{bmatrix}$ and writing out the rows of $\mathcal{H}_{n-1}$ as row vectors $\mathbf{d}_i$:

$$\text{diag}\left(\begin{bmatrix} \mathbf{y}^u \\ \mathbf{y}^d \end{bmatrix}\right) \begin{bmatrix} \mathbf{d}_1 & \mathbf{d}_1 \\ \vdots & \vdots \\ \mathbf{d}_{2^{n-1}} & \mathbf{d}_{2^{n-1}} \\ \mathbf{d}_1 & -\mathbf{d}_1 \\ \vdots & \vdots \\ \mathbf{d}_{2^{n-1}} & -\mathbf{d}_{2^{n-1}} \end{bmatrix} \begin{bmatrix} \mathbf{u} \\ \mathbf{v} \end{bmatrix} > \mathbf{0} \qquad \text{Eq. 2}$$

Now, we construct two matrices $\mathbf{F}$ and $\mathbf{G}$ according to whether the function output flips sign when the last variable $x_n$ flips sign. With the ordering adopted for the rows of $\mathcal{H}_n$, this corresponds to comparing $y_i^u$ with $y_i^d$ for $i = 1, 2..2^{n-1}$. With slight abuse of notation, we can write $\mathbf{F} = \{y_i^u \mathbf{d}_i | y_i^u = y_i^d\}$ and $\mathbf{G} = \{-y_i^d \mathbf{d}_i | y_i^u \neq y_i^d\}$. Thus $\mathbf{F}$ and $\mathbf{G}$ are made up of disjoint rows of $\mathcal{H}_{n-1}$ which may have been scaled by $-1$. It can be shown that the original inequality system can be written in terms of $\mathbf{F}$ and $\mathbf{G}$ as follows [16]:

$$\mathbf{Fu} > \mathbf{Fv} > -\mathbf{Fu}$$
$$\mathbf{Gv} > \mathbf{Gu} > -\mathbf{Gv}$$

Further, this system can be converted to an equivalent system for $\mathbf{u} - \mathbf{v}$ and $\mathbf{u} + \mathbf{v}$:

$$\begin{bmatrix} \mathbf{F} \\ -\mathbf{G} \end{bmatrix}(\mathbf{u} - \mathbf{v}) > \mathbf{0} \qquad \begin{bmatrix} \mathbf{F} \\ \mathbf{G} \end{bmatrix}(\mathbf{u} + \mathbf{v}) > \mathbf{0}$$

The solution space of these inequality systems can be conveniently found since $\begin{bmatrix} \mathbf{F} \\ -\mathbf{G} \end{bmatrix}$ and $\begin{bmatrix} \mathbf{F} \\ \mathbf{G} \end{bmatrix}$ are almost Sylvester type Hadamard matrices, where some rows may be negated and/or permuted with respect to our natural row order. So, we can apply Theorem 1 to write the expressions for $\mathbf{u} - \mathbf{v}$ and $\mathbf{u} + \mathbf{v}$, since row permutations does not change the fact that all the solutions are spanned by positive vectors from $\mathbb{R}^{2^{n-1}}$. Having the expressions for $\mathbf{u} - \mathbf{v}$ and $\mathbf{u} + \mathbf{v}$ as functions of arbitrary positive vectors (and rows of $\mathbf{F}$ and $\mathbf{G}$), we can eventually write $\mathbf{u}$ and $\mathbf{v}$ individually. This in turn, gives the full solution space for the original inequality in terms of positive (row) vectors $\boldsymbol{\alpha}, \boldsymbol{\alpha}', \boldsymbol{\gamma}, \boldsymbol{\gamma}'$ of appropriate size as follows [16]:

$$\mathbf{w}^T = \begin{bmatrix} \mathbf{u}^T \\ \mathbf{v}^T \end{bmatrix} = \begin{bmatrix} \boldsymbol{\alpha} + \boldsymbol{\alpha}' & -\boldsymbol{\gamma} + \boldsymbol{\gamma}' \\ -\boldsymbol{\alpha} + \boldsymbol{\alpha}' & \boldsymbol{\gamma} + \boldsymbol{\gamma}' \end{bmatrix} \begin{bmatrix} \mathbf{F} \\ \mathbf{G} \end{bmatrix} = \quad \boldsymbol{\alpha}, \boldsymbol{\alpha}', \boldsymbol{\gamma}, \boldsymbol{\gamma}' > \mathbf{0} \qquad \text{Eq. 3}$$

From this expression it can be shown that either the $\mathbf{u}$ part or the $\mathbf{v}$ part of the solution can be forced to have $2^{n-2}$ zeros as outlined next.

The theorem proceeds symmetrically according to whether the number of rows in $\mathbf{F}$, ($\#\mathbf{F}$) is larger than the number of rows in $\mathbf{G}$, ($\#\mathbf{G}$). Let's assume $\#\mathbf{G} \geq \#\mathbf{F}$, then $\#\mathbf{G}$ elements of $\mathbf{u}$ part of the solution vector $\mathbf{w}$ can be zeroed by arbitrarily setting $\boldsymbol{\alpha}, \boldsymbol{\alpha}'$ and computing $\boldsymbol{\gamma}, \boldsymbol{\gamma}'$ accordingly. To see this, note that $\mathbf{G}$ is of full rank, and thus $-\boldsymbol{\gamma} + \boldsymbol{\gamma}'$ can be used to generate arbitrary real numbers at $\#G$ component positions. These positions can be identified by finding a full rank column subset of $G$ by, for example row Echelon reduction. So, $\#G$ components of $u$ can be forced to become zero since $u = (\alpha + \alpha')F + (-\gamma + \gamma')G$. The argument for $\#F > \#G$ is proven symmetrically (exchange the roles of $\alpha, \alpha'$ with $\gamma, \gamma'$). Therefore, since $\#G + \#F = 2^{n-1}$ we can always zero at least $2^{n-2}$ components of $w$. □

## 5. Weight upper bound on PTFs with length at most $0.75 \times 2^n$

**Theorem 3**. *Any n-variable BF can be represented by a PTF with length at most $0.75 \times 2^n$ and integer weights bounded above by $2^{n2^{n-3} - 2^{n-1} + 4n}$. In particular, by using the order $h = 2^n$, of the Hamdard matrix involved, we have*

$$\lceil w \rceil \leq 2^{0.125h \log h - 0.5h + 4 \log h} \qquad Eq.\ 4$$

To prove this theorem, we first prove a statement about determinants of the submatrices of Hadamard matrices.

**Lemma 6**. *Given a Hadamard matrix $\mathcal{H}$ in block form $\mathcal{H} = \begin{bmatrix} F_1 & F_2 \\ G_1 & G_2 \end{bmatrix}$, $G_1$ is invertible if and only if $F_2$ is invertible.*

*Proof.* Assume $G_1$ is invertible. Take an arbitrary vector $a_1 \in \mathbb{R}^{\#F}$. Clearly there exists a vector $a_2 \in \mathbb{R}^{\#G}$ such that $[a_1^T \ a_2^T]\begin{bmatrix} F_1 \\ G_1 \end{bmatrix} = 0$ (i.e. take $a_2^T = -a_1^T F_1 G_1^{-1}$). This means that $\begin{bmatrix} a_1 \\ a_2 \end{bmatrix}$ is a linear combination of the columns of $\begin{bmatrix} F_2 \\ G_2 \end{bmatrix}$ because $\mathcal{H}$ is orthogonal. Therefore, there exists $x \in \mathbb{R}^{\#F}$ with $\begin{bmatrix} a_1 \\ a_2 \end{bmatrix} = \begin{bmatrix} F_2 \\ G_2 \end{bmatrix} x$. In particular, this means that for arbitrary $a_1 \in \mathbb{R}^{\#F}$ there is always a linear combination $z$ of the columns of $F_2$ such that $a_1 = F_2 x$, i.e. $F_2$ spans $\mathbb{R}^{\#F}$. Thus $F_2$ is of full rank. Since it is square, it must be invertible. To see the double implication, replace the roles $\begin{bmatrix} F_1 \\ G_1 \end{bmatrix}$ with $\begin{bmatrix} F_2 \\ G_2 \end{bmatrix}$ and $a_1$ with $a_2$. □

*Corollary.* Consider a submatrix $A$ of a Hadamard matrix $\mathcal{H}$ obtained by deleting the $k$ rows indexed by $i_R \in R$ and $k$ columns index by $j_C \in C$, where R and C are k-sized subsets of $\{1, 2, \ldots, n\}$. Then, $A$ is invertible if and only if the matrix formed by the intersection of the deleted rows and columns, i.e. $[\mathcal{H}_{i_R, j_C}]$ is invertible.

*Proof.* By pre- and post- multiplying $\mathcal{H}_n$ with appropriate permutation matrices $A$ can be brought to the lower left of the permuted matrix taking the role of $G_1$ whereas $[\mathcal{H}_{i_R, j_C}]$ is taking the role of $F_2$. Since the row and column permutation of a Hadamard matrix is still Hadamard, the corollary follows. □

### *Proof of Theorem 3*.

By Theorem 2 we know that a given n-variable BF induces two matrices $F$ and $G$ with $2^{n-1}$ columns. Let $f = \#F$ and $g = \#G$, i.e. the number of rows in $F$ and $G$. These matrices allows us to construct a polynomial sign-representation with at most $0.75 \times 2^n$ monomials.

In fact, all sign representing polynomials (i.e. their coefficient vectors $w$) are determined by arbitrarily chosen positive (appropriately sized) row vectors $\alpha, \alpha', \gamma, \gamma'$ (Eq. 3). According to Theorem 2, if $g \geq f$ [$f > g$], we know that $g$ [$f$] elements of $u$ [$v$] part of the coefficient vector can be zeroed by arbitrarily selecting $\alpha, \alpha'$ [$\gamma, \gamma'$] and appropriately computing $\gamma, \gamma'[\alpha, \alpha']$ based on the selection. Without loss of generality let's assume that $g \geq f$, and further assume that the elements that will be zeroed are the first $g$ elements so that we can we can write $F$ and $G$ as $[F_1 \; F_2]$ and $[G_1 \; G_2]$ respectively. Then Eq. 3 becomes:

$$\begin{bmatrix} \alpha + \alpha' & -\gamma + \gamma' \\ -\alpha + \alpha' & \gamma + \gamma' \end{bmatrix} \begin{bmatrix} F_1 & F_2 \\ G_1 & G_2 \end{bmatrix} = \begin{bmatrix} u^T \\ v^T \end{bmatrix} \quad \alpha, \alpha', \gamma, \gamma' > 0 \qquad Eq.\ 5$$

Since $G_1$ is invertible, for any $\alpha, \alpha' > 0$ we can always find $\gamma, \gamma' > 0$ such that

$$[\alpha + \alpha' \quad -\gamma + \gamma'] \begin{bmatrix} F_1 \\ G_1 \end{bmatrix} = 0 \qquad Eq.\ 6$$

This means that the vector $[\alpha + \alpha' \quad -\gamma + \gamma']^T$ is a linear combination of the columns of $\begin{bmatrix} F_2 \\ G_2 \end{bmatrix}$ since $\begin{bmatrix} F_1 & F_2 \\ G_1 & G_2 \end{bmatrix}$ is orthogonal. Therefore, there exist a real vector $x$ such that

$$[\alpha + \alpha' \quad -\gamma + \gamma']^T = \begin{bmatrix} F_2 \\ G_2 \end{bmatrix} x \qquad Eq.\ 7$$

Conversely, for any $x$ satisfying $F_2 x > 0$ we can find $\alpha, \alpha', \gamma, \gamma' > 0$ such that Eq. 6 holds and thus generate a sign representation with at most $0.75 \times 2^n$ monomials with

$$\alpha + \alpha' = x^T F_2^T \text{ and } -\gamma + \gamma' = x^T G_2^T.$$

In the above derivation, observe that we can freely choose to require $\alpha = \alpha'$. Substituting $\alpha + \alpha'$ and $-\gamma + \gamma'$ in the original solution (Eq. 5) with $\alpha = \alpha'$ choice, we obtain the following expression for the coefficient vector $w$:

$$w^T = \begin{bmatrix} u^T \\ v^T \end{bmatrix} = \begin{bmatrix} x^T(F_2^T F_1 + G_2^T G_1) & x^T(F_2^T F_2 + G_2^T G_2) \\ (\gamma + \gamma')G_1 & (\gamma + \gamma')G_2 \end{bmatrix} \qquad Eq.\ 8$$

Since $\begin{bmatrix} F_1 & F_2 \\ G_1 & G_2 \end{bmatrix}$ is orthogonal we see that

$$w^T = \begin{bmatrix} u^T \\ v^T \end{bmatrix} = \begin{bmatrix} 0 & 2^{n-1} x^T \\ (\gamma + \gamma')G_1 & (\gamma + \gamma')G_2 \end{bmatrix} \qquad Eq.\ 9$$

Now let's find bounds for $[x]$ and $[\gamma + \gamma']$ in the following lemmas.

***Lemma 7.*** *(upper bound of $[x]$). $[x]$ is upper bounded by* $2^{n-2^{n-2}}(2^{n-2} - 1)^{\frac{2^{n-2}-1}{2}}$

*Proof.* Let's consider the case, $g \geq f$ as assumed in the main theorem. From Lemma 6 we know that $F_2$ is invertible. Hence for any $\alpha + \alpha'$ there is some $x$ satisfying $F_2 x = [\alpha + \alpha']^T$. Choose $\alpha_j = \alpha'_j = 0.5r$ so that we have the system of linear equations $F_2 x = r\mathbf{1}$ where $r$ is a positive constant integer to be substituted later. Following the technique of Håstad [18], the solution $x$ to this equation can be found by Cramer's rule. Accordingly, the solution to $Ax = b$ is given by $x_j = \frac{|A_j|}{|A|}$ where $A_j$ is the matrix obtained by taking $A$ and replacing the j$^{th}$ column with $b$. In our case, we have $b = r1$ and $A = F_2$. Since $A$ is a $\pm 1$ matrix, according to Hadamard's inequality (Lemma 4), $|A| \leq (f)^{\frac{f}{2}}$. Writing the determinant $|A_j|$ by using the cofactor expansion along the replaced column we have:

$$|A_j| = \sum_{i=1}^{f} b_i (-1)^{i+j} M_{i,j} \qquad \text{Eq. 10}$$

Where $M_{i,j}$ corresponds to the (i, j) minor, i.e. the determinant of $A_j$ where j$^{th}$ the column and i$^{th}$ row are deleted. So, $M_{i,j}$ is the determinant of a $(f-1) \times (f-1)$ matrix with entries $\pm 1$. Thus $|A_j| \leq rf(f-1)^{\frac{f-1}{2}}$. On the other hand, according to Lemma 5, $|A|$ and $|A_j|$ are integers that are divisible by $2^{f-1}$ and $2^{f-2}$ respectively. Therefore by letting $r=|A|/2^{f-2}$ we are guaranteed that the $x_j = \frac{|A_j|}{|A|}$ will be integer and upper bounded by $2^{2-f} f(f-1)^{\frac{f-1}{2}}$ in magnitude i.e.

$$[x] \leq 2^{2-f} f(f-1)^{\frac{f-1}{2}} \qquad \text{Eq. 11}$$

It is easy to show that when $f > g$ this becomes

$$[x] \leq 2^{2-g} g(g-1)^{\frac{g-1}{2}} \qquad \text{Eq. 12}$$

Combining both case yields

$$[x] \leq 2^{n-2^{n-2}} (2^{n-2} - 1)^{\frac{2^{n-2}-1}{2}} \qquad \text{Eq. 13}$$

**Lemma 8.** *(upper bound on $[w]$). $[w]$ is upper bounded by $3 \times 2^{n-2}[x]$.*

*Proof.* Let's find upper bounds for the upper and lower halves of the solution vector $w$ separately:

i. $u$ part:

Since $u$ is given by $u^T = [0 \quad 2^{n-1} x^T]$ (Eq. 9) it is clearly integer and bounded by $2^{n-1}[x]$, i.e. $[u] \leq 2^{n-1}[x]$

ii. $v$ part:

Since $v = (\gamma + \gamma')[G_1 \quad G_2]$ (Eq. 9), we seek an upper bound for $[\gamma + \gamma']$. We know that $[-\gamma + \gamma']^T = G_2 x$ must hold; but we are free to choose $\gamma, \gamma' > 0$ to satisfy this. Let $\beta = G_2 x$ for brevity and choose $\gamma, \gamma'$ as follows:

$$\gamma_i = \begin{cases} -\beta_i + 0.5, & \text{if } \beta_i < 0 \\ 0.5, & \text{else} \end{cases} \qquad \gamma'_i = \begin{cases} 0.5, & \text{if } \beta_i < 0 \\ \beta_i + 0.5, & \text{else} \end{cases}$$

Observe that the required equality $[-\pmb{\gamma}+\pmb{\gamma}']^T = \pmb{G_2}\pmb{x}$ is satisfied and the components of $\pmb{\gamma}+\pmb{\gamma}'$ are positive integers when $\pmb{x}$ is integer. Furthermore, due to the construction of $\pmb{\gamma},\pmb{\gamma}'$ we have $\lceil\pmb{\gamma}+\pmb{\gamma}'\rceil = \lceil\pmb{G_2}\pmb{x}\rceil + 1$, which at once yields

$$\lceil\pmb{v}\rceil \leq (\lceil\pmb{G_2}\pmb{x}\rceil + 1)\mathcal{g}$$

where $\mathcal{g} = \#\pmb{G_1} = \#\pmb{G_2}$. Noting that $\lceil\pmb{G_2}\pmb{x}\rceil \leq \lceil\pmb{x}\rceil\mathcal{f}$ we get an expression for an upper bound on $\pmb{v}$:

$$\lceil\pmb{v}\rceil \leq \lceil\pmb{x}\rceil\mathcal{g}\mathcal{f} + \mathcal{g}\ .$$

Combining (i) and (ii) we establish an upper bound for the coefficients:

$$\lceil\pmb{w}\rceil \leq \max\{\mathcal{g}\mathcal{f}+\mathcal{g}, 2^{n-1}\}\lceil\pmb{x}\rceil \qquad\qquad Eq.\ 14$$

It is possible to show that when $\mathcal{f} > \mathcal{g}$ this becomes

$$\lceil\pmb{w}\rceil \leq \max\{\mathcal{g}\mathcal{f}+\mathcal{f}, 2^{n-1}\}\lceil\pmb{x}\rceil \qquad\qquad Eq.\ 15$$

Since $\mathcal{f} \leq 2^{n-2}$ [$\mathcal{g} \leq 2^{n-2}$] and $\mathcal{f}+\mathcal{g} = 2^{n-1}$ we have $\lceil\pmb{w}\rceil \leq (2^{2(n-2)} + 2^{n-2})\lceil\pmb{x}\rceil$, so assuming n>1 we can write

$$\lceil\pmb{w}\rceil \leq 2^{4n-3}\lceil\pmb{x}\rceil \qquad\qquad Eq.\ 16$$

□

Finally, combining the last two results given in Eq. 13 Eq. 11 and Eq. 16 we get

$$\lceil\pmb{w}\rceil \leq 2^{4n-1-2^{n-2}}(2^{n-2}-1)^{\frac{2^{n-2}-1}{2}}$$

replacing $2^{n-2}-1$ with $2^{n-2}$ for simplicity we get

$$\lceil\pmb{w}\rceil \leq 2^{n2^{n-3}-2^{n-1}+4n} \qquad\qquad Eq.\ 17$$

Substituting $h = 2^n$ we reach the desired result:

$$\lceil\pmb{w}\rceil \leq 2^{0.125h\log h - 0.5h + 4\log h} \qquad\qquad Eq.\ 18$$

□

***Corollary to Theorem 3***. The PTF weight of n-variable Boolean functions with PTF representations with $0.75 \times 2^n$ monomials are bounded above by $0.75x2^{n2^{n-3}-2^{n-1}+5n}$.

$$\mathcal{W}_f[\mathcal{D}_f \leq 0.75 \times 2^n] \leq 0.75 \times 2^{n2^{n-3}-2^{n-1}+5n} \qquad\qquad Eq.\ 19$$

*Proof.* Since we have at most $0.75 \times 2^n$ coefficients, we multiply the right-hand size of Eq. 18 with $0.75 \times 2^n$ to get to the given upper bound. □

## 6. Bent functions with short PTFs and small weights

Bent functions are interesting set of BFs introduced by Rothaus [19], which have complicated combinatorial properties with significance in cryptanalysis [20, 21]. In this report, we are interested in their PTF representation in terms of weight and length. Semi-bent function are another related class of functions with important cryptographic properties [22], which we also use in this section. We first give the definitions of these function classes.

*Definition* (Bent function). An n-variable Boolean function is called bent if and only if the Walsh coefficients of the function are all $\pm 2^{n/2}$.

*Definition* (Semi-bent function). If the Walsh spectrum of an n-variable Boolean function $f$ takes only the values 0 or $\pm 2^{(n+1)/2}$ then $f$ is a semi-bent function.

*Remark*. Bent functions exist in only even dimensions; whereas semi-bent functions exist only in odd dimension.

**Lemma 9.** *Let $f_b$ be an n-variable bent function and b be its Walsh polynomial, then it can be decomposed into two semi-bent functions $f_p$ and $f_q$ with Walsh polynomials p and q with the following relation:*

$$b(x_1, x_2 \cdots, x_n) = 0.5(x_n + 1)p(x_1, x_2 \cdots, x_{n-1}) + 0.5(-x_n + 1)q(x_1, x_2 \cdots, x_{n-1})$$

*Proof.* It is clear that if polynomials $p$ and $q$ part are Walsh polynomials for $f_p$ and $f_q$ then $b$ is a Walsh polynomial for $f_b$ so let's look at the semi-bentness. Let $[u^T \ v^T]$ the coefficients of $b$. Note that as $b(x_1, x_2 \cdots, 1) = p(x_1, x_2 \cdots, x_{n-1})$ and $b(x_1, x_2 \cdots, -1) = q(x_1, x_2 \cdots, x_{n-1})$, the upper half and lower halves of $f_b$ coincides with $f_p$ and $f_q$. Applying Lemma 2 and the Eq. 1 we have:

$$\begin{bmatrix} u \\ v \end{bmatrix} = \begin{bmatrix} \mathcal{H}_{n-1} & \mathcal{H}_{n-1} \\ \mathcal{H}_{n-1} & -\mathcal{H}_{n-1} \end{bmatrix} \begin{bmatrix} f_p \\ f_q \end{bmatrix}$$

By multiplying both sides by $[I \ \ I]$ and $[I \ \ -I]$ we get

$$\mathcal{H}_{n-1} f_p = (u+v)/2$$
$$\mathcal{H}_{n-1} f_q = (u-v)/2$$

Note that the left-hand sides are exactly the (Walsh) coefficients of $p$ and $q$ (due to Lemma 2). Further, observe that they can only take values of 0 or $\pm 2^{n/2}$ since the components of $u$ and $v$ are $\pm 2^{n/2}$. Thus, $f_p$ and $f_q$ are semi-bent functions of order $n-1$. □

**Lemma 10.** *An n-variable semi-bent function have $2^{n-1}$ number of zeros in its Walsh spectrum.*

*Proof.* Let $k$ be the number of non-zero Walsh coefficients of an n-variable semi-bent function. Then, the norm of the Walsh spectrum will be $k^{1/2} 2^{(n+1)/2}$ which must be equal to $2^n$, as the Walsh spectrum of all n-variable BFs is $2^n$ (due to Lemma 3). Solving $k^{1/2} 2^{(n+1)/2} = 2^n$ we see that the number of non-zero coefficients is $k = 2^{n-1}$. So, the number of zeros is also $2^{n-1}$ since we have $2^n$ coefficients in total. □

**Theorem 4.** *Any n-variable bent function can be represented as a polynomial threshold functions by using $0.75 \times 2^n$ monomials with integer coefficients less than or equal to $2^n$ in absolute value.*

To prove this, let's first prove a Lemma concerning the inner workings of Theorem 2.

**_Lemma 11_**. *If $f$ is a bent function then the row-sum of $\begin{bmatrix} F \\ G \end{bmatrix}$ and $\begin{bmatrix} F \\ -G \end{bmatrix}$ have $2^{n-2}$ zeros, where up $F$ and $G$ are the matrices formed by the application of Theorem 2 to $f$.*

*Proof.* Write $f$ as partitioned into upper and lower halves so that $f^T = \begin{bmatrix} f_{up}^T & f_{lo}^T \end{bmatrix}$, and observe that the rows making up $F$ and $G$ matrices, are unique rows of a Sylvester type Hadamard matrix of order $n-1$ which are scaled by $f_{up,i}$ for some $i \in \{1, 2, \ldots, 2^{n-1}\}$. Thus $\begin{bmatrix} F \\ G \end{bmatrix}$ is a Hadamard matrix which is almost Sylvester type, except that some rows might be permuted and/or negated due to the scaling by $f_{up,i}$. Luckily, we can recover the natural order by appropriate row permutations, and further we can pull out the scaling factor outside to obtain a Sylvester type Hadamard matrix. The same arguments are also true for $\begin{bmatrix} F \\ -G \end{bmatrix}$, where this time the scaling is done with the lower half $f$. Thus, for some permutation matrices $P_{up}$ and $P_{lo}$, we have the following relations:

$$\text{diag}(f_{up}^T)\mathcal{H}_{n-1} = P_{up}\begin{bmatrix} F \\ G \end{bmatrix}$$

$$\text{diag}(f_{lo}^T)\mathcal{H}_{n-1} = P_{lo}\begin{bmatrix} F \\ -G \end{bmatrix}$$

Now, taking the row-sum of both sides of the equations (i.e. multiplying by $\mathbf{1}^T$), we see that left sides become the Walsh coefficients of the functions $f_{up}^T$ and $f_{lo}^T$, which are semi-bent according to Lemma 9. Also noting that $\mathbf{1}^T M = \mathbf{1}^T P M$ for any arbitrary matrix $M$ and any permutation matrix $P$ we have:

$$\varpi_{up}^T = \mathbf{1}^T \begin{bmatrix} F \\ G \end{bmatrix} \text{ and } \varpi_{lo}^T = \mathbf{1}^T \begin{bmatrix} F \\ -G \end{bmatrix}$$

Since the Walsh coefficients $\varpi_{up}^T$ and $\varpi_{lo}^T$ of semi-bent functions have $2^{n-2}$ zeros due to Lemma 10, the proof of the Lemma is complete. □

**_Proof of Theorem 4_**. From Theorem 2, we know that for a given BF $f$, once $F$ and $G$ matrices are constructed, any $\alpha, \alpha', \gamma, \gamma' > 0$ assignment produces a polynomial threshold function representation of $f$ with the weights $w$ given below (replicated from Eq. 5)

$$w^T = \begin{bmatrix} u^T \\ v^T \end{bmatrix} = \begin{bmatrix} \alpha + \alpha' & -\gamma + \gamma' \\ -\alpha + \alpha' & \gamma + \gamma' \end{bmatrix} \begin{bmatrix} F \\ G \end{bmatrix} \quad \alpha, \alpha', \gamma, \gamma' > 0 \qquad Eq.\ 20$$

To show that at least one fourth of the elements of $w$ can be always zeroed, Theorem 3 creates zeros in either $u$ or the $v$ part of the solution by finding a full rank square submatrix (by row Echelon reduction) in $G$ or $F$, and use its inverse to eliminate some of the terms appearing in $u$ or $v$. In the case of Bent functions, we are lucky: we neither need to apply Echelon reduction to detect the elements to be zeroed, nor need to take inverse. By using Lemma 11 we can choose $\alpha, \alpha', \gamma, \gamma'$ in a straightforward manner to readily get $2^{n-2}$ zeros in the coefficient vector $w$. To be concrete, let $\alpha = \alpha' = 0.5 \times \mathbf{1}$ and $\gamma = 0.5 \times \mathbf{1}$, $\gamma' = 1.5 \times \mathbf{1}$ so that $u^T = \mathbf{1}^T \begin{bmatrix} F \\ G \end{bmatrix}$ which has $2^{n-2}$ zeros according to Lemma 11. Another alternative is to let $\alpha = \alpha' = 0.5 \times \mathbf{1}$ and $\gamma = 1.5 \times \mathbf{1}$, $\gamma' = 0.5 \times \mathbf{1}$. In which case we would have $u^T = \mathbf{1}^T \begin{bmatrix} F \\ -G \end{bmatrix}$, which again would have $2^{n-2}$ zeros. Thus, due to Lemma 11, in either case we are guaranteed to have at least $2^{n-2}$ zero coefficients in the polynomial threshold representation of $f$. So, with the aforementioned choice of $\alpha, \alpha', \gamma, \gamma'$ we obtain the same density bound that is given in Theorem 2 for general BFs. However, now, we can greatly improve the weight upper bound with this choice of parameters. To see this, plug in the proposed choices of $\alpha, \alpha', \gamma, \gamma'$ in Eq. 20 to get:

$$w^T = \begin{bmatrix} 1 & \pm 1 \\ 0 & 2 \times \mathbf{1} \end{bmatrix} \begin{bmatrix} F \\ G \end{bmatrix} \qquad \text{Eq. 21}$$

Which immediately tells us that $\lceil w \rceil \leq 2^n$. □

***Corollary to Theorem 4.*** The PTF weight of n-variable Bent functions with PTF representations with at most $0.75 \times 2^n$ monomials are bounded above by $0.75 \times 2^{2n}$.

$$\mathcal{W}_{f_{bent}}[\mathcal{D}_{f_{bent}} \leq 0.75 \times 2^n] \leq 0.75 \times 2^{2n} \qquad \text{Eq. 22}$$

*Proof.* Each coefficient is upper bounded by $2^n$ and we have at most $0.75 \times 2^n$ of those coefficients, thus the result follows. □

## 7. PTF density and weight of sparse Boolean functions

In this section, we address PTF representation of BF that are constant except for some small number of variable assignments. Let's define this concretely.

*Definition*. A BF $f$ is called m-sparse, if $m = \min(|f^{-1}(1)|, |f^{-1}(-1)|) \ll 2^n$

An upper bound on the PTF degree of m-sparse BF functions was obtained as $\lfloor \log m \rfloor + 1$ in a recent repot [10]; however, to our knowledge there is no PTF density bound better than the known for general BFs, i.e. $0.75 \times 2^n$ [16]. Here we show that m-sparse BF can always be represented with $m + 2^{n-1}$ monomials for small $m$. We state this as a theorem.

***Theorem 5.*** Let $f$ be an m-sparse BF with $m \leq 2^{n-1}$, then $f$ has a PTF representation with length $\leq 2^{n-1} + \min(m, 2^{n-2})$

Proof. To prove this claim, we can start from a constant BF $f$, say $f = 1$ and assess the effect of transforming $f$ into an m-sparse function on $F$ and $G$ matrices produced through the application of Theorem 2.

Note when $m \geq 2^{n-2}$ the theorem statement is already satisfied by Theorem 2. So, let's consider the case $m < 2^{n-2}$ and recall the sign-representation condition, and write $f$ as partitioned into upper and lower halves:

$$\text{diag}\left(\begin{bmatrix} f^u \\ f^d \end{bmatrix}\right) \begin{bmatrix} \mathcal{H}_{n-1} & \mathcal{H}_{n-1} \\ \mathcal{H}_{n-1} & -\mathcal{H}_{n-1} \end{bmatrix} w > 0$$

When we consider the constant function $f = 1$ [$f = -1$], $f^u = f^d = \mathbf{1}$ [$f^u = f^d = -\mathbf{1}$], and thus according to the construction in Theorem 2, we will have $\mathbf{F} = \mathcal{H}_{n-1}$ and $\mathbf{G} = [\,]$. Now to convert this constant function to an m-sparse function, one needs to pick m components of $f = \begin{bmatrix} f^u \\ f^d \end{bmatrix}$ and negate them. There are $\binom{2^n}{m}$ possibilities, each of which generate certain $\mathbf{F}$ and $\mathbf{G}$ matrices. For our purposes, only the size of them are relevant, as according to Theorem 2, the number of zeros obtained in the sign representation is at least $\max(\#\mathbf{F}, 2^{n-1} - \#\mathbf{F})$. Let's calls this quantity $q(\#\mathbf{F})$. Remember that $\#\mathbf{F}$ is equal to the number of matches between the upper and the lower half of $f$, i.e. $\#\mathbf{F} = |\{i | f_i^u = f_i^d\}|$ and is equal to $2^{n-1}$ when we start with the constant $f$. Thus, we can imagine an adversarial negation strategy to change $f$ into an m-sparse matrix to minimize $q(\#\mathbf{F})$, thereby minimizing the number of zeros in the final solution. Since $2^{n-1} \geq q(\#\mathbf{F}) \geq 2^{n-2}$ what the adversarial choice can do is to reduce $\#\mathbf{F}$ from $2^{n-1}$ as much as possible without increasing $2^{n-1} - \#\mathbf{F}$ beyond $2^{n-2}$. The latter case is not possible since $m < 2^{n-2}$ so the adversarial choice does not need to worry about it. Thus, the adversarial choice minimizes $q(\#\mathbf{F})$ by creating $m$ upper-lower mismatches (i.e. apply m negations such that $m = |\{i | f_i^u \neq f_i^d\}|$). This strategy yields $q(\#\mathbf{F}) = 2^{n-1} - m$ zeros which is the minimum possible for m-negations. Thus, we conclude that $q(\#\mathbf{F}) \geq 2^{n-1} - m$ for any m-sparse function. Therefore, for any m-sparse function it is possible to construct a PTF representation with $2^{n-1} + m$ or a smaller number of monomials. □

**Theorem 6**. *Let f be an m-sparse BF with $m \leq 2^{n-1}$, then f has a integer-coefficient PTF representation with length $\leq 2^{n-1} + \min(m, 2^{n-2})$ with coefficient magnitudes $\leq 2^{n+0.5m \log m - m + 1.5 \log m + 2}$.*

*Proof.* Note in our m-sparse construction $\mathit{f} \geq \mathit{g}$ where $\mathit{f} = \#\mathbf{F}$ and $\mathit{g} = \#\mathbf{G}$. From Lemma 7 and Lemma 8 (Eq. 12 and Eq. 15) we know that for a general BF (with $\mathit{f} \geq \mathit{g}$), an upper bound on the absolute value of the PTF weights, $w$ is given by
$$[w] \leq \max\{\mathit{g}\mathit{f} + \mathit{f}, 2^{n-1}\}[x]$$
where
$$[x] \leq 2^{2-\mathit{g}}\mathit{g}(\mathit{g} - 1)^{\frac{\mathit{g}-1}{2}}$$
So, to obtain a weight upper bound for the m-sparse function case, what remains to be done is to plug in the upper bounds of $\mathit{g}\mathit{f} + \mathit{f}$ and $\mathit{g}$ for m-sparse functions, which are attained at $\mathit{f} = 2^{n-1} - m$ and $\mathit{g} = m$. So, we have:

(i) $[w] \leq \mathit{g}\mathit{f} + \mathit{f} = (2^{n-1} - m)(m + 1)[x]$

(ii) $[x] \leq 2^{2-m} m^{\frac{m+1}{2}}$

Where to get (ii) we substituted $m$ instead of $m - 1$ for $\mathit{g} - 1$ in the base of the power expression. Now, plugging (ii) in (i) and taking log of both sides we have
$$\log[w] \leq \log(2^{n-1} - m) + \log(m + 1) + (2 - m) + (0.5m + 0.5)\log m$$
For the sake of a compact expression, we remove $-m$ from the argument of first log, and replace 1 with $m$ in the argument of the second log to obtain:
$$\log[w] \leq n + 0.5m \log m - m + 1.5 \log m + 2$$

Raising both sides to the power 2 we get
$$[w] \leq 2^{n+0.5m \log m - m + 1.5 \log m + 2}$$
□

**Corollary to Theorem 6.** *For any of n-variable m-sparse Boolean function f with $m \leq 2^{n-2}$ there exists a PTF representation with length at most $m + 2^{n-1}$ and PTF weight bounded by $3 \times 2^{2n + 0.5m \log m - m + 1.5 \log m}$. In other words, we have:*

$$\mathcal{W}_{f_{sp}}\left[\mathcal{D}_{f_{sp}} \leq m + 2^{n-1}\right] \leq 3 \times 2^{2n + 0.5m \log m - m + 1.5 \log m}$$

*Proof.* Multiply the upper bound found for $[w]$ by $0.75 \times 2^n$ as it is an upper bound on the number of non-zero coefficients in $w$. □

## 8. Conclusion

In an earlier report we have shown that any BF can be represented as a PTF with at most $0.75 \times 2^n$ monomials (i.e. nonzero weight input lines) [16]. However, no upper bound on the absolute value of the coefficients was known when the coefficients were constrained to integers. In this report, we fill this gap by establishing an upper bound on the PTF weight of general Boolean functions when they are represented with at most $0.75 \times 2^n$ monomials. In addition, we study m-sparse and bent BFs. For the former we obtain new density and weight bounds, which indicate that low-m sparse BFs can be represented with low density and low weight. For the bent functions, we show that they assume a surprisingly low weight PTF representation when they are represented with $0.75 \times 2^n$ monomials. When the PTF representation of a bent function with its Walsh coefficient is considered, the number of monomials involved is $2^n$ and the absolute value of the coefficients are $2^{n/2}$. When we wish to sign-represent a bent function without using $2^{n-2}$ monomials (*which*, depends on the function) then the coefficient magnitudes increase; but, merely become at most $2^n$. We think that this bound is tight; however, the bound established for general BF is far from being tight. Thus, future work is needed to tighten the weight bound for general BFs.

## 9. Acknowledgements

Support for this work is provided by the International Joint Research Promotion Program, Osaka University under the project "Developmentally and biologically realistic modeling of perspective invariant action understanding".## 10. References

[1] Rumelhart DE, Hinton GE, Williams RJ. Learning internal representations by error propagation. In: Rumelhart DE, McClelland JL, group aP, editors. Parallel Distributed Processing1986. p. 151-93.
[2] Giles CL, Maxwell T. Learning, invariance, and generalization in high-order neural networks. Apllied Optics. 1987;26:4972-8.15